## FUTURAMA, MARVEL'S SUPER-VILLAINS AND BOLTZMANN BRAINS

Mario Daniel Martín – The Australian National University
https://researchers.anu.edu.au/researchers/martin-md

## Abstract

In this article I will explore some uses of the concept of Boltzmann brains in comics (Marvel's The Incredible Hercules), an animated sitcom (Futurama) and a film (Guardians of the Galaxy Vol. 2). Previous similar uses of the concept (before Boltzmann brains were named as such) will be traced, and potential influences on these artistic outlets will be analysed. I will also explain how the concept was used in a novel published in 2018 by Ediciones Ayarmanot in Buenos Aires ("La inevitable resurrección de los cerebros de Boltzmann", to be published in English as "Iridio Einnui vs. the Boltzmann brains").

After briefly reviewing the second law of thermodynamics and the concept of Boltzmann brains, the paper analyses four instances in which it has been used for artistic purposes, answering two key questions:

a) How entities who (in theory) appear in the foam of interstellar space and suddenly disappear are presented in the fictional scenario?

b) How the plot justifies that Boltzmann brains could exist for a time long enough to allow them to become viable fictional characters?

This is a slightly extended version of a paper to be published in the proceedings of the Science Fiction Convention *Pórtico*, Universidad Tecnológica Nacional, Sede Regional Córdoba, Argentina.

## FUTURAMA, LOS SUPERVILLANOS DEL UNIVERSO MARVEL Y LOS CEREBROS DE BOLTZMANN

## Ejes temáticos:

La ciencia ficción en el cine y la televisión, en los comics, música, fotografía, etc.

Ciencia Ficción y edición – Novedades editoriales.

## Resumen

En este trabajo voy a explorar los usos que se le ha dado a un concepto abstracto de la cosmología, los cerebros de Boltzmann, en diversos medios, concentrándome en las historietas, los dibujos animados y una película. También resumiré cómo he usado ese mismo concepto en una novela publicada en 2018 por Ediciones Ayarmanot.

**Palabras clave:** Cerebros de Boltzmann, Futurama, Guardianes de la Galaxia 2, El increíble Hércules, La inevitable resurrección de los cerebros de Boltzmann

## 1. La segunda ley de la termodinámica

Las leyes de la termodinámica se formularon durante la revolución industrial para tratar de explicar cómo se comportaba la energía en las máquinas de vapor. La primera ley o la ley de la conservación de la energía, relaciona el trabajo y el calor transferido intercambiado en un sistema cerrado usando el concepto de energía, que, en la formulación clásica de la ley ni se crea ni se destruye, sino se transforma. En el caso de las máquinas de vapor, el calor se transforma en trabajo mecánico.

La segunda ley nos permite entender cómo los procesos naturales evolucionan en el tiempo. En el caso de la máquina de vapor, se estableció empíricamente, al explorar la eficiencia de una máquina térmica, que no es posible un proceso que convierta todo el calor emitido en trabajo. Más tarde, se estableció como un principio usando el concepto abstracto de la entropía, que tradicionalmente se considera una medida del desorden de un sistema, y se extrapoló más allá de las máquinas térmicas, a todo el universo. La ley, en su formulación clásica, dice que la entropía del universo tiende a un máximo, es decir que siempre aumenta. Lo que genera la pregunta ¿por qué observamos un universo con baja entropía, es decir un universo ordenado? El científico austríaco Ludwig Boltzmann propuso que esto podría deberse a una fluctuación aleatoria en un universo de alta entropía. Es decir, vivimos en una burbuja de baja entropía, en un inmenso universo donde la mayoría de las otras regiones tienen alta entropía [1-2]. Boltzmann definió la entropía en forma estadística, no como una ley en donde siempre se incrementa por definición. Crucialmente, explica la entropía como el logaritmo del número de microestados equivalentes de un sistema, y si la entropía se incrementa, es porque los estados "desordenados" son más probables. Pero eso también permite que la entropía se reduzca, con menor probabilidad. Eso nos permite volver a la pregunta de por qué observamos un universo con baja entropía. Si el universo es lo suficientemente grande, pueden existir este tipo de regiones, simplemente por las leyes de la probabilidad [3-4]. Para entender el razonamiento de Boltzmann, imaginemos que abrimos una botella de perfume en una habitación cerrada. El perfume se evaporará, y se difundirá por el aire de la habitación. Las moléculas se distribuirán en forma aleatoria, en miles de configuraciones. Ahora bien, si esperamos el tiempo suficiente, y eso es un tiempo larguísimo, va a haber eventualmente una reconfiguración de las moléculas tal que todas vuelvan a estar en el frasco de perfume. La probabilidad de que eso suceda es infinitamente pequeña, pero si esperamos suficiente tiempo, en algún momento, por las leyes de la probabilidad, va a suceder. Este razonamiento, aparentemente inverosímil, es una consecuencia de que Boltzmann redefinió el concepto de entropía en términos estadísticos. Al hacerlo, invocando los átomos, que entonces eran un concepto dudoso, se adelantó a su tiempo. Y las críticas de otros científicos lo llevaron al suicidio. La historia de la ciencia, sin embargo, lo reivindicó con creces.

## 2. La expansión acelerada del universo y los cerebros de Boltzmann

A fines del siglo XX los astrónomos descubrieron que la expansión del universo, en vez de desacelerarse como predeciría la acción de la fuerza de la gravedad, en realidad se aceleraba [5]. Eso reavivó especulaciones cosmológicas sobre el tamaño del universo (más allá del universo observable), y generó que se revisara la hipótesis propuesta por Boltzmann. En un universo infinito (que además dura infinitamente), aún los eventos más improbables pueden suceder. En particular, esta revisión generó la llamada paradoja de los cerebros de Boltzmann. Éstas son entidades hipotéticas conscientes de sí mismas, que aparecen en el espacio sideral a causa de una fluctuación estadística de la materia, y desaparecen después de un corto tiempo [6-8]. De la misma forma en que Boltzmann propuso que nuestro universo podría ser el producto de una fluctuación, dichas entidades, nombradas en su honor, son más probables que la creación de una burbuja que contenga galaxias. Esto es simplemente una consecuencia de las leyes de la estadística, cuando se aplican a grandes números de casos. La probabilidad de una fluctuación está determinada por el volumen de la materia que se "ordena" aleatoriamente. Entonces es más probable que se cree un planeta que un sistema solar, y un sistema solar que una galaxia. Y por el mismo tipo de razonamiento, es más probable que se cree un cerebro que un ser humano (o un extraterrestre) completo, y esto más probable que espontáneamente aparezca un planeta en donde se desarrollen observadores evolutivos (como los seres humanos). La consecuencia de este tipo de especulación es que, a largo plazo, teorías cosmológicas que permiten estas fluctuaciones van a predecir que en el universo los observadores más típicos, o más numerosos, serían estos efímeros y cognitivamente inestables cerebros de Boltzmann [9-10].

La principal aplicación que se le ha dado al concepto de los cerebros de Boltzmann es usarlo como una típica "reducción al absurdo" para discriminar entre distintas teorías cosmológicas. Aquellas que predicen que son los observadores más frecuentes, son dudosas [11]. Pero a medida que el concepto se fue haciendo más popular y fue discutido en las revistas de divulgación científica, empezó a influenciar otras áreas, especialmente porque se vincularon a teorías cosmológicas que postulaban multiversos [12] o universos paralelos [13-14], y a la teoría de cuerdas [15]. Otras discusiones, consideradas "esotéricas" por los científicos tradicionales, como postulados que el universo en realidad está compuesto por matemáticas, también se vincularon a los cerebros de Boltzmann [16]. Los filósofos de la ciencia también se involucraron en el debate, algunos para concluir que no hay forma de decidir si uno es o no un cerebro de Boltzmann [17], otros para negar ese postulado [18], y otros para proponer que el concepto permitía una avenida a la inmortalidad [19-20] (ya que si todo lo que es posible puede suceder en un universo infinito, es posible que la mente del lector que está leyendo esto en este momento pueda ser reproducida por azar en algún lugar, en algún momento). Los teólogos también intervinieron en los numerosos debates cruzados [21], principalmente para atacar a las teorías que proponían multiversos como una explicación de por qué el universo era propicio para la vida (sin intervención de deidades).

Por supuesto, dada la popularidad y lo inverosímil de los postulados, también la cultura popular y el arte ha usado el concepto. En las siguientes secciones exploraremos cómo se reflejan los desarrollos esbozados anteriormente en los cómics, los dibujos animados, las películas y la literatura. Como el tema es muy amplio, nos limitaremos a explorar dos preguntas:

a) ¿Cómo se usan en la ficción entidades que aparecen en medio del espacio sideral y luego desaparecen tan pronto como han nacido?

b) En particular, ¿cómo el argumento de la obra de ficción justifica que puedan existir por un tiempo suficientemente largo para ser personajes viables?

## 3. Futurama

La serie de ciencia ficción de dibujos animados Futurama tiene varios episodios en donde presuntos cerebros de Boltzmann son personajes significativos. No son nombrados como tales, pero son cerebros que flotan en el espacio exterior, y, como discutiremos más adelante, se han relacionado con los cerebros de Boltzmann. En este artículo nos concentraremos en dos de los episodios, que nos permiten saber cómo los cerebros voladores son trasladados a la ficción: *El día que la Tierra se quedó estúpida* (episodio 39, o episodio 7 de la temporada 3) [22] y *El porqué de Fry* (episodio 64, o episodio 10 de la temporada 4) [23].

La serie sigue las desventuras de Philip J. Fry, un joven repartidor de pizzas que ha sido congelado criogénicamente por accidente durante mil años, y despierta en el año 3000 en una futurística Nueva York. Durante toda la serie trata de conquistar a Leela, una joven cíclope. A través de la intervención de su único descendiente vivo, el Profesor Hubert J. Farnsworth, se convierte en un repartidor de la compañía Planet Express, fundada por su tataratarara…sobrino para distribuir mercancías en los planetas cercanos a la Tierra. Leela es la capitana de la nave de mensajería y carga estelar. Otros personajes importantes son el robot alcohólico Bender Bending Rodríguez (fabricado en México) y Mordelón, una mascota extraterrestre de Leela, un nibloniano que puede devorar animales mucho más grandes que él en pocos segundos. Los excrementos de Mordelón están compuestos de "materia oscura" y son muy valiosos porque pueden servir como combustible para las naves espaciales.

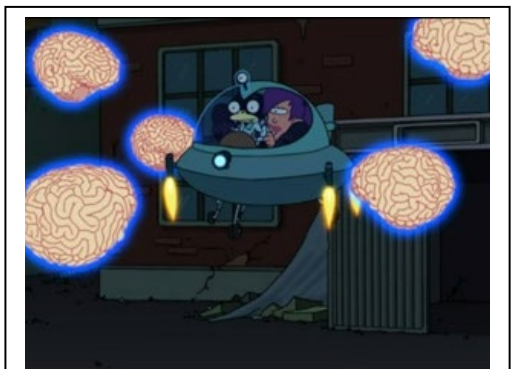

Figura 1. Mordelón escapa con Leela de la tierra invadida por los cerebros flotantes

La primera aparición de los cerebros flotantes es en el episodio *El día que la Tierra se quedó estúpida*. Los cerebros invaden la tierra y todos se vuelven estúpidos, excepto Fry. Leela es transportada por Mordelón al planeta de los niblonianos, que está en el centro exacto del universo, donde se le explica que cuando se formó el universo en el big bang, los niblonianos ya existían. Un milisegundo después del big bang se crearon los engendros cerebrales, y desde entonces ha habido una guerra entre los niblonianos y los cerebros, que están comandados por un cerebro inmenso. Mordelón pretende ser una inocente mascota, que no habla, pero en realidad es el embajador nibloniano en la tierra.

Los niblonianos le explican a Leela que los cerebros odian a todos los seres conscientes, porque el pensamiento los exaspera. Viajan por el universo convirtiendo a todos los seres inteligentes en estúpidos para eliminar el pensamiento. Están comandados por el Gran Cerebro, un cerebro gigante que "tiene un centro viscoso lleno de odio". Leela quiere saber por qué Fry es inmune. La razón es porque su cerebro no tiene ondas cerebrales delta. Los engendros cerebrales interfieren con las ondas delta de los seres conscientes. En la tierra, los humanos, los animales, los robots y algunas plantas generan esas ondas. Como Fry es inmune, es la única esperanza de salvar al universo.

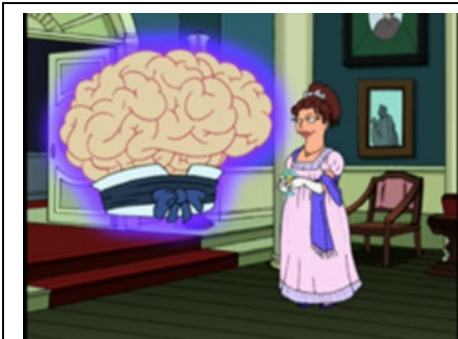

Figura 2. El Gran Cerebro dentro del mundo ficcional de *Orgullo y Prejuicio* de Jane Austen

El episodio concluye con una lucha entre Fry y el Gran Cerebro. Para detener a Fry, el Gran Cerebro lo hace entrar en distintos mundos paralelos. Son mundos de la ficción (*Moby Dick, Las aventuras de Tom Sawyer*, y

*Orgullo y Prejuicio*). Dentro de la ficción, Fry escribe su propia novela, llena de errores de ortografía y tramas inconclusas. Pero en la novela hace que el Gran Cerebro abandone la tierra por una "rason" inexplicable. Eso les permite a los niblonianos comer a los otros cerebros más pequeños. Cuando todo vuelve a la normalidad, los habitantes de la tierra no recuerdan nada de lo sucedido y no le creen a Fry cuando les cuenta que él salvó al planeta. Como Leela no recuerda tampoco lo que pasó, Mordelón vuelve a pretender ser su inocente mascota.

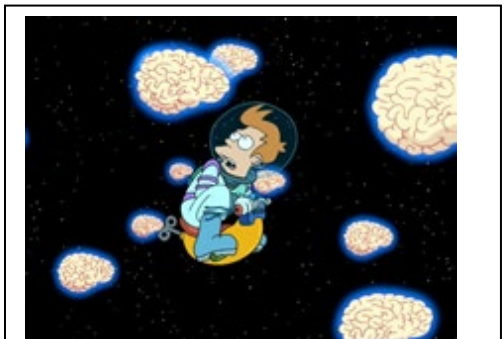

Figura 3. Fry viajando hacia el lugar donde está la Infoesfera.

Los engendros cerebrales vuelven a aparecer en el episodio *El porqué de Fry.* Mordelón lleva a Fry a su planeta, y allí le encargan la misión de destruir la Infoesfera, una base de datos con todo el conocimiento del universo que han recopilado los cerebros. Lo que pretenden es completar la base de datos y luego destruir el universo para que no aparezca nueva información. Fry descubre que la razón por la que su cerebro no genera ondas delta, y por lo tanto es inmune a la influencia de los cerebros flotantes, se debe a que es su propio abuelo. Eso es el resultado de haberse acostado con su abuela cuando viajó al pasado en el episodio *Todo anda bien en Roswell* (episodio 51, o episodio 19 de la temporada 3) [24], lo que es una sátira de *Volver al futuro II*, y también de la paradoja del abuelo [25]. Los niblonianos le dan una bomba de interfaz cuántica que debe poner dentro de la Infoesfera para enviarla a un universo alternativo para siempre y una motonave espacial a cuerda (un scooter) que lo lleva al

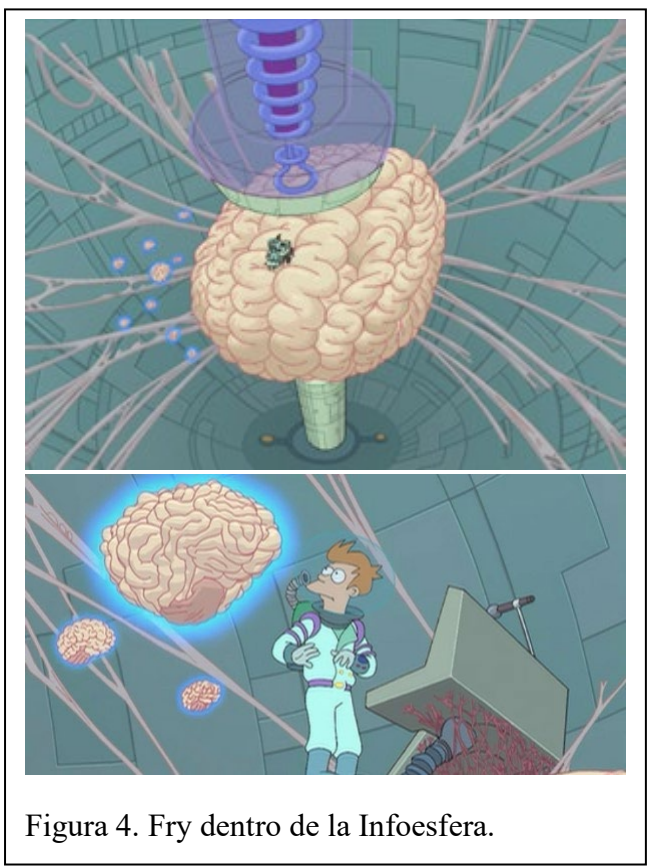

Figura 4. Fry dentro de la Infoesfera.

lugar. Le dice que no debe pensar muy intensamente para no ser detectado.

Fry logra eludir a los cerebros centinelas que custodian la Infoesfera, y entrar a su interior, donde ahora reside el Gran Cerebro. Pero en vez de activar la bomba de inmediato y escapar se entretiene porque se da cuenta de que puede encontrar la respuesta a todas las preguntas del universo en el lugar. Descubre, por ejemplo, que la goma de las estampillas postales está hecha con moco de sapo, y que los cerebros flotantes fueron la causa de la extinción de los dinosaurios. El Gran Cerebro descubre que Fry está ahí, e intenta distraerlo. Su scooter se rompe, pero lo mismo decide sacrificarse por el bien del universo. Uno de los cerebros flotantes le sugiere que averigüe lo que pasó el día en que fue congelado criogénicamente. Descubre que Mordelón lo empujó en la máquina que lo congeló, o sea, que no fue enviado al futuro accidentalmente, como él creía. Los cerebros flotantes logran enviarlo al pasado, y aparece una copia suya debajo de la mesa donde Mordelón estaba escondido antes de empujarlo. Puede, en teoría, cambiar el pasado y detener a Mordelón, pero éste le dice que es la única persona en el universo que podía ser reclutado para detener a los cerebros flotantes, y por eso decidieron enviarlo al futuro. Fry decide no detener su congelamiento por su amor a Leela, y se empuja a sí mismo dentro de la máquina que lo crioniza. Es enviado adentro de la Infoesfera otra vez, y puede escapar porque Mordelón, que ya sabe lo que puede suceder, le da un mejor scooter. Vuelve a la tierra con Mordelón, y éste le hace olvidar todo lo que sucedió en su viaje.

Hay una gran cantidad de información en internet sobre los episodios de Futurama. Sin embargo, en los wikis y otros sitios de seguidores de la serie hay pocas referencias a los cerebros de Boltzmann, incluso en las páginas dedicadas a estos episodios, o a los cerebros voladores [26-28]. Cuando se explora las referencias y las parodias en los episodios considerados, se cita una película de terror llamada *El cerebro del Planeta Arous* [29]. Es muy difícil discernir si la literatura científica y los artículos en la prensa de divulgación sobre los cerebros de Boltzmann, que empezaron a discutirse en la época en que los episodios fueron escritos, han tenido influencia. Sin embargo, es sabido que la serie tiene numerosas referencias a la física y las matemáticas, y los que escribieron los guiones tienen doctorados en ciencias exactas [30-31]. Dada esta formación científica, es posible que hayan sentido nombrar a los cerebros de Boltzmann, especialmente porque Martin Rees, en su libro *Antes del Comienzo* especula sobre estas entidades potenciales en el momento en que la serie estaba siendo escrita, ya que, de acuerdo al físico Alexander Vilenkin, influyó en el desarrollo de dicho concepto. [32]

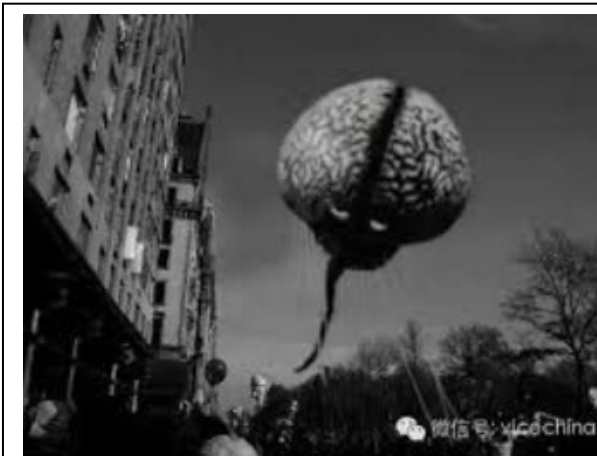

Figura 5. Una imagen de la película *El cerebro del planeta Arous*.

Volviendo a nuestra pregunta sobre cómo se hace que persistan en el tiempo, la serie no lo explicita. Asume que los cerebros fueron creados en el big bang, y persistieron desde entonces. O sea, no son los clásicos efímeros cerebros de Boltzmann que desaparecen después de un corto tiempo. El hecho de que no hayan respetado su idiosincrasia, a pesar de la formación científica de los guionistas, puede deberse a que desconocían la paradoja. Alternativamente, la causa puede deberse a lo que David Cohen, el coordinador de la serie, explicó en una entrevista dedicada a las referencias matemáticas y científicas en la serie:

> De vez en cuando, incorporábamos referencias bastante oscuras a la ciencia, siempre y cuando no distrajeran a los espectadores promedios… Teníamos la esperanza de que, a pesar de que ese material pasaría desapercibido para la mayoría, convertiría en seguidores apasionados de la serie a aquellos que pudieran apreciarlo [31a]

En otras palabras, la comedia era más importante que la ciencia.

## 3. Guardianes de la galaxia, volumen 2.

En esta película uno de los personajes, llamado Ego, se originó como un cerebro de Boltzmann. Cuando el personaje explica su origen en el guión original del director de James Gunn [33], vemos, en el texto subrayado, que nace como un cerebro de Boltzmann clásico:

> *Esta forma que ves frente a ti es sólo una extensión de quién soy en realidad. <u>No sé de dónde vine, exactamente. La primera cosa que recuerdo es centellear a la deriva en el cosmos, completamente solo.</u> Me alimenté de la materia alrededor mío, como el plankton. Me hice más fuerte y más inteligente. Formé una covertura para protegerme de la intemperie. Y continué creando a partir de ahí, capa a capa, el planeta mismo en el que ahora caminas. Construí las espirales que se extienden hasta el cielo y los tuneles ahondando en sus profundidades. Pero no era una hormiga satisfecha unicamente por su labor. Necesitaba algo más. Deseaba… tener un significado. Debe haber otra vida en el universo, aparte de mí, pensé, y me propuse encontrarla.*

La imagen que acompaña a esta explicación puede verse en la figura 6.

Sería muy complejo detallar todo lo que sucede después con este personaje en la película. Un buen resumen se encuentra en la entrada de Wikipedia sobre Ego [34]:

> *Sin embargo, al estar aburrido de la inmortalidad y decepcionado con un universo lleno de vida inferior, Ego decide rehacer todos los mundos del universo en extensiones de sí mismo, un plan que requiere plantas de semillero plantadas y el poder de otro Celestial para activarlas. Para lograr esto, Ego se aparea con varias especies hasta que se conciba* una descendencia adecuada para ayudar a poner en práctica su plan. *El único niño adecuado resulta ser Peter Quill, aunque se rebela y eventualmente ayuda a destruir a Ego cuando Quill se entera de que Ego mató a su madre Meredith Quill.*

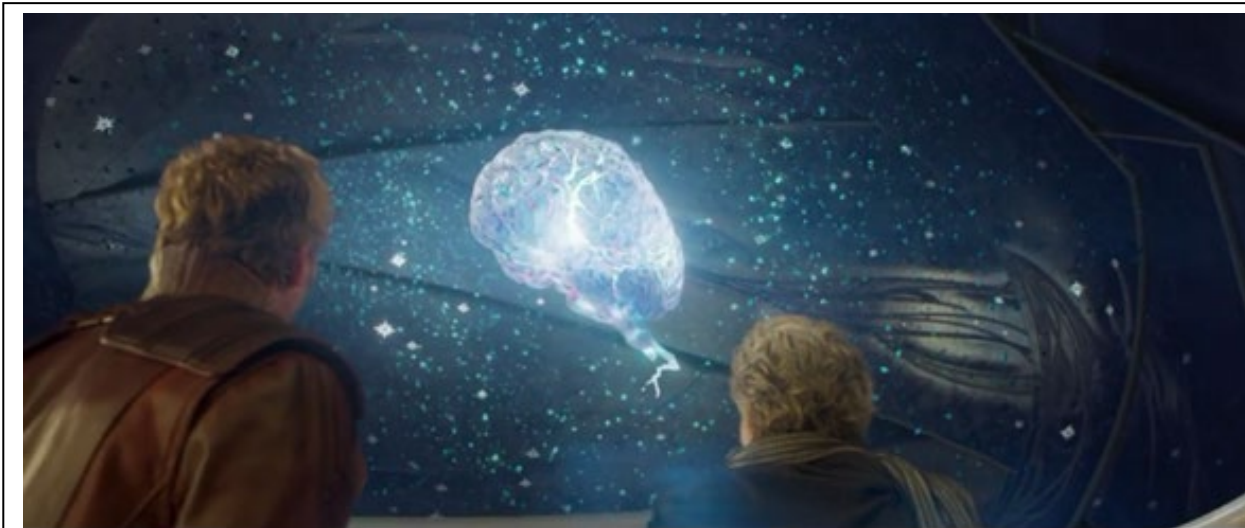

Figura 6. Escena que acompaña a la explicación del origen de Ego en *Guardianes de la Galaxia 2*.

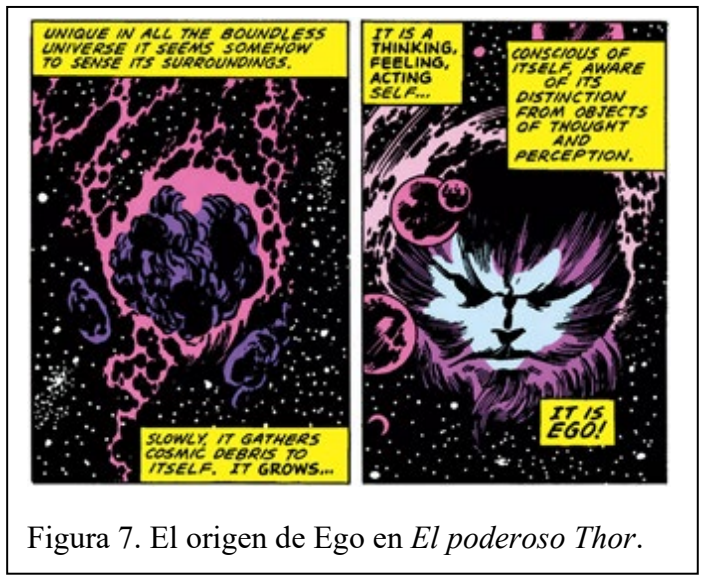

Figura 7. El origen de Ego en *El poderoso Thor*.

Ego es un supervillano clásico. Pero también es un planeta viviente. El hecho que se origine como un cerebro de Boltzmann es algo que se agrega en esta película, porque Ego tiene una larga historia en las historietas de Marvel antes de ser reciclado por Gunn. Aparece por primera vez en *El poderoso Thor* número 132 en 1966, pero es recién en la segunda serie de apariciones en la historieta (números 201 a 228) que se conoce su origen, y, como puede verse en la figura 7, se formó como todos los planetas, acumulando material disperso en el cosmos gracias a la fuerza de gravedad. Lo único inusual es que es un planeta con conciencia. El origen de esta "vida" no se explica. En otra serie, *Los Cuatro Fantásticos versus Ego* en 1981, aparece como un cerebro flotando en el aire, pero en realidad los antagonistas están en el centro del planeta Ego, y llegan a su cerebro para neutralizar su maldad.

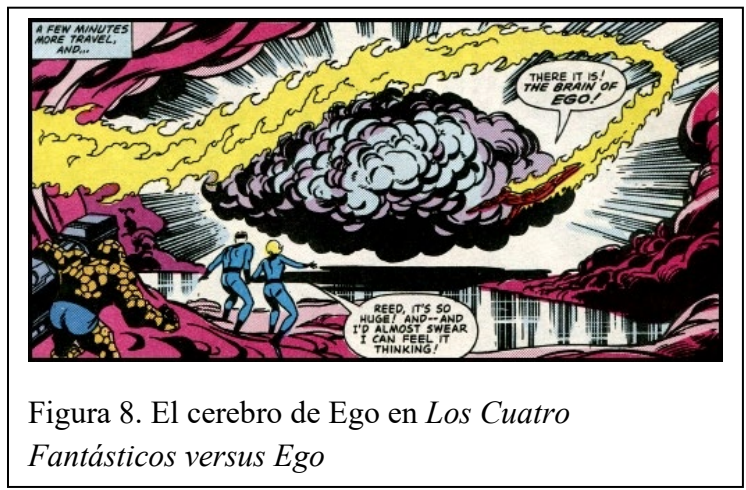

Figura 8. El cerebro de Ego en *Los Cuatro Fantásticos versus Ego*

Podría hacerse una larga cronología sobre la influencia de los descubrimientos astronómicos sobre Ego, ya que el personaje se recicla en muchas otras series más. Podría también escribirse una cronología de la influencia del psicoanálisis en las historietas de Marvel, porque hay otro planeta viviente llamado Super-ego, que luego se convierte en un Alter-ego que gira alrededor de Ego, que posteriormente es un Ello (en inglés Id), una luna egoísta que gira alrededor de Ego, y luego se independiza para ejercer su maldad. Sin embargo, para nuestros propósitos, la explicación de por qué el cerebro de Boltzmann presentado en la película *Guardianes de la galaxia, volumen 2* persiste en el tiempo y no desparece después de experimentar su conciencia en el espacio sideral se encuentra en la "prehistoria" del personaje, en las otras historietas, a la que el guionista hace referencia en la película. Presentarlo como un cerebro de Boltzmann es simplemente una "actualización" de la astronomía presupuesta en la ficción para justificar su origen. Y no va más allá. Es, como antes, la semilla a partir de la que se aglutina la materia para formar un planeta, y, gracias a esta actualización, se explica por qué es consciente. El Ego "actualizado" está claramente influenciado por el concepto de los cerebros de Boltzmann, pero estas entidades no son nombradas explícitamente, ni su rol o sus características son elaboradas en profundidad. Dado esto, no es necesario explicar explícitamente en la película por qué persiste en el tiempo.

## 5. El increíble Hércules

En esta serie de historietas encontramos una referencia explícita a los cerebros de Boltzmann, y por eso la analizaremos en más detalle.

La serie de Marvel Comics es una continuación de *El Increíble Hulk*, que llega hasta el número 111. A partir del número 112, cambia el nombre a *El Increíble Hércules*, y se centra en otro personaje, Hércules, que es el héroe griego, ahora transformado en superhéroe del panteón norteamericano, que pelea junto a otros superhéroes modernos, por ejemplo, el Capitán América, Los Vengadores, y el Hombre Araña [35].

Un personaje importante en esta serie es Amadeus Cho, un joven coreano-norteamericano, un héroe predominantemente intelectual [36-37]. En realidad, Amadeus es un personaje muy complejo, y, como en el caso de Ego, apareció en varias series de Marvel anteriores. Y se transforma luego, en otras series creadas después de *El Increíble Hércules*, una nueva versión de *El Increíble Hulk*, llamado *El Totalmente Impresionante Hulk*, un héroe al mismo tiempo intelectual y con una prodigiosa fuerza física.

Los capítulos 133, 135 y 137 de *El Increíble Hércules* presentan la historia de Amadeus Cho, y son las que nos interesan porque contienen referencias a los cerebros de Boltzmann. Esas historietas nos permiten descubrir la historia de Amadeus, un niño prodigio. Amadeus es el hijo de Helen y Phil Cho, emigrantes coreanos a EEUU, quienes le dieron ese nombre en homenaje a Mozart. Nació en Tucson, Arizona, y sus padres se dieron cuenta rápidamente que era un genio [38]. Cuando Amadeus tenía 15 años, participó en un concurso llamado *Batalla Cerebral* (Brain Fight) organizado por una compañía de jabones llamada Excello, donde ganó el primer premio de 5.000 dólares y fue declarado "la séptima persona más inteligente del mundo".

En realidad, el concurso había sido organizado por un científico paralítico, un antiguo niño prodigio llamado Pythagoras Dupree [39], cuyos propósitos eran identificar a posibles competidores suyos en términos de inteligencia, para eliminarlos. Es el equivalente "intelectual" de la madrastra de Blancanieves. Pythagoras se había autoproclamado la "sexta persona más inteligente del mundo", y no se sabe quiénes eran los otros cinco antes de él. Sin embargo, Amadeus era lo suficientemente inteligente para ser un desafío para Phytagoras, y por ello, envía a sus hombres para destruir la casa de los Cho. La casa explota, y mueren los padres, pero Amadeus se salva. Los hombres enviados por Pythagoras lo persiguen, y es así como conoce a Bruce Banner (El increíble Hulk), quien lo protege de sus perseguidores. También se inicia una amistad y asociación mágica entre Amadeus y Hércules.

En el episodio 135 un adolescente Amadeus viaja a la ciudad de Excello, donde debería haber recibido su premio cuando niño. Allí es asistido por el agente Sexton, una detective del FBI. En la ciudad encuentran a personas que no son reales, porque aparentemente son cerebros de Boltzmann disfrazados de humanos. Inicialmente creen que son agentes del Doctor Japanazi, un científico con dos cerebros, autoproclamado miembro del eje del mal (los enemigos de los aliados en la segunda guerra mundial), pero luego se entiende que Japanazi es un avatar de Dupree. De hecho, "japanazi" es un término derogatorio en inglés norteamericano con el que se nombraba a los miembros del eje. Su inclusión se debe a la historia de Phytagoras Dupree, que creció leyendo historietas del universo Marvel, en la época en que se desarrollaba la segunda guerra mundial, cuando gozaban de inmensa popularidad.

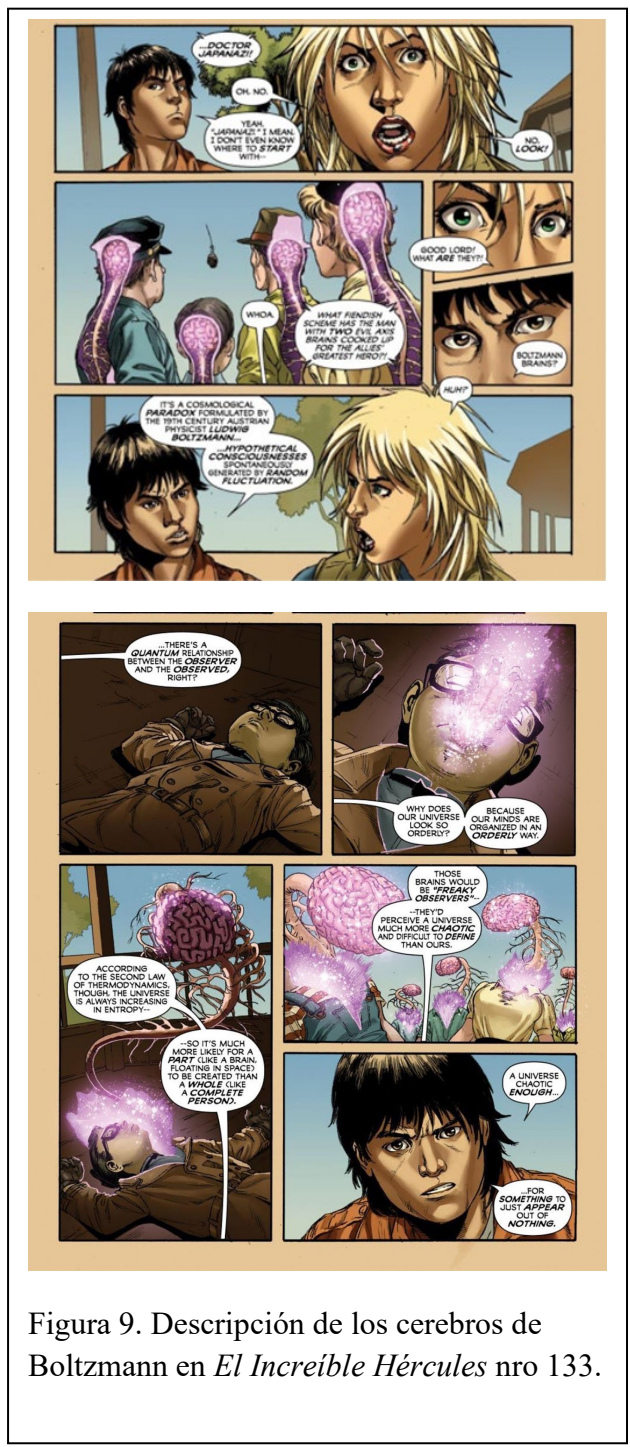

Figura 9. Descripción de los cerebros de Boltzmann en *El Increíble Hércules* nro 133.

Amadeus le explica a la agente Sexton qué son los cerebros de Boltzmann. Vale la pena traducir el diálogo (que puede verse en el original en inglés en la figura 9):

-Dios Santo, ¿qué son?

-Cerebros de Boltzmann. Es una paradoja cosmológica formulada por el científico austríaco Ludwig Boltzmann en el siglo XIX. Son conciencias hipotéticas generadas por fluctuaciones aleatorias.

- ¿Qué?

-Hay una relación cuántica entre el observador y lo observado, ¿verdad? ¿Por qué el universo nos parece tan ordenado? Porque nuestras mentes están organizadas de una manera ordenada. De acuerdo con la segunda ley de la termodinámica, sin embargo, la entropía del universo siempre está incrementándose. Esos cerebros de Boltzmann serían "observadores raros", porque perciben un universo mucho más caótico y difícil de definir que el nuestro. Es un universo lo suficientemente caótico para que algo aparezca de la nada.

El personaje del niño del que sale un cerebro de Boltzmann en la segunda sección de la figura es en realidad una personificación de Phytagoras Dupree, el antagonista de Amadeus como un niño.

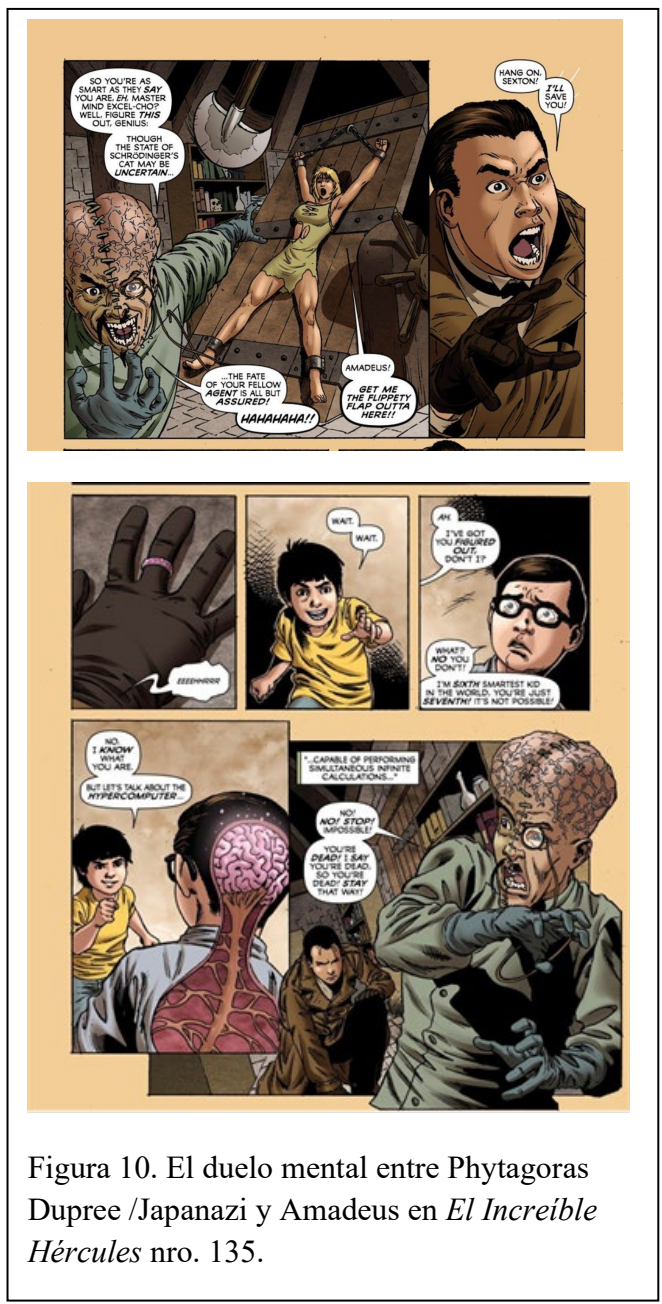

Figura 10. El duelo mental entre Phytagoras Dupree /Japanazi y Amadeus en *El Increíble Hércules* nro. 135.

En la próxima entrega de la serie que continúa la historia encontramos a un adulto Amadeus que enfrenta al Doctor Japanazi en su laboratorio, quien ha capturado a la agente Sexton. En el laboratorio hay una caja que contiene un gato que está simultáneamente vivo y muerto, una alusión al famoso experimento mental de Erin Schrödinger.

En paralelo, vemos una disputa entre Amadeus y Phytagoras Dupree, que aparece de nuevo convertido en un niño, como puede verse en la figura 10. Allí nos enteramos de que todas las escenas, incluyendo las que sucedieron en la ciudad que se han descrito anteriormente, son una especie de duelo mental entre Amadeus y Phytagoras, quien ha creado esos universos paralelos para matar a Amadeus. En esta entrega de la serie encontramos otro diálogo entre Amadeus y la agente Sexton, que nos aclara un poco más el rol de los cerebros de Boltzmann en la trama:

-*Tus cerebros de Boltzmann otra vez.*

-*Sí, su presencia es lo que está creando ésta, y las otras realidades. Ellos están "observando" lo que Dupree ha programado en ellos, reflejándose a sí mismos, básicamente.*

Más adelante, después de que Japanazi es ajusticiado por la agente Sexton, y descubren que se encuentran en una sala que está a punto de explotar, el diálogo, mostrado en la figura 11, nos permite saber más sobre como los cerebros de Boltzmann son concebidos en la historieta.

*-Espera, no es el momento de entrar en pánico.*

*-No es pánico, es una parálisis momentánea, inducida por una desesperación total.*

*-Escucha, desde el momento en que entramos a la ciudad de Excello, Dupree nos ha atrapado en* un laberinto de "realidades burbujas" que él ha creado, congelándonos en una multiplicidad de estados cuánticos, como el gato que estaba atrapado en la caja del Doctor Japanazi.

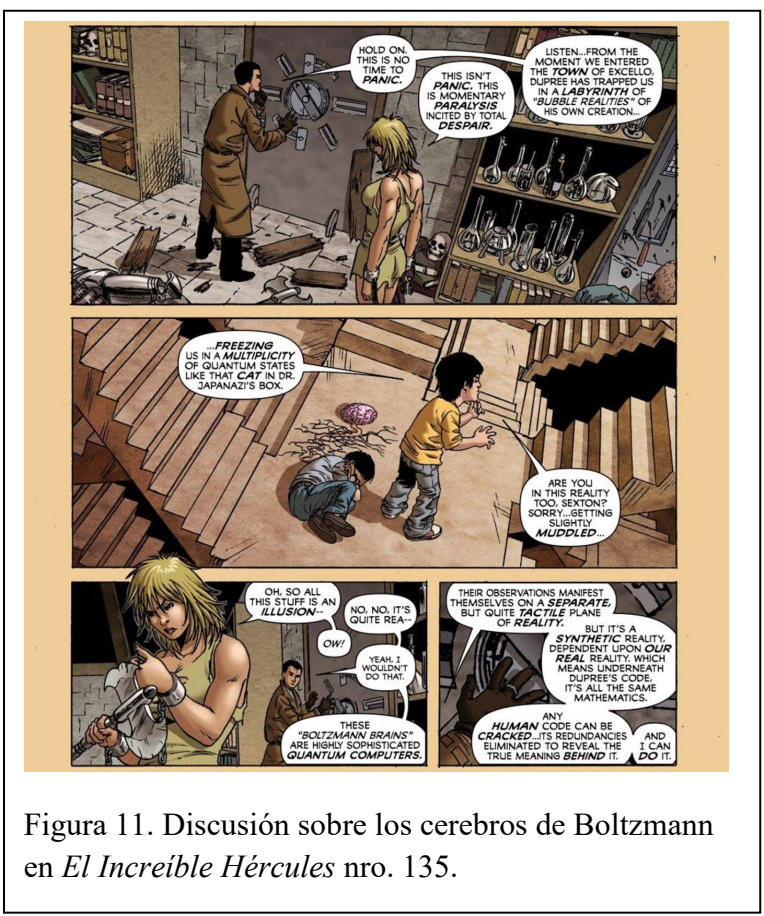

Figura 11. Discusión sobre los cerebros de Boltzmann en *El Increíble Hércules* nro. 135.

Aquí la escena cambia a mostrar nuevamente al niño Amadeus, que ha vencido al niño Dupree, quién está caído en el piso en la viñeta al centro de la figura 11. También hay un cerebro de Boltzmann caído en el piso, y muchas escaleras que potencialmente llevan a la salida de esa realidad en donde peleaban Amadeus y Dupree.

*- ¿Estás también tú en esta realidad,*

*Sexton? Disculpa, está todo borroso.*

*-Oh, entonces esto es una ilusión…*

*-No, no. Es bastante real.*

- *¡Guau!*

*-Esos cerebros de Boltzmann son computadoras cuánticas bastante sofisticadas. Sus observaciones se manifiestan en ámbito separado de la realidad, que es bastante táctil. Pero es una realidad sintética, que depende de nuestra realidad real, lo que significa que debajo del código de Dupree es la misma matemática. Todo código humano puede ser descifrado, y sus redundancias eliminadas para revelar el verdadero significado detrás de él. Y yo puedo descifrarlo.*

Lo que nos aclara este último dialogo es que los cerebros son las baterías de los avatares mentales creados por Dupree, quien los ha programado. No tienen otro rol que proveer la energía para esas burbujas / realidades y, en apariencia, se limitan a observarse / observar. No son agentes en la historia, porque son marionetas mentales de Dupree. Lo que importa, para el argumento, es que las dos mentes "hiper-computadoras" de Amadeus y Dupree están peleando, para determinar cuál de los universos paralelos prevalecerá. Las numerosas escaleras (a la Escher) que se ven en la viñeta en donde Amadeus es un niño en la figura 11 corresponden a las distintas rutas para salir de ese multiverso (de acuerdo a la interpretación de Everett de la mecánica cuántica) creado usando la "energía" de los cerebros de Boltzmann. Al final del episodio, Amadeus logra hacer colapsar la función de onda del universo en la realidad "real" y descubre que Dupree está escondido en un edificio en el medio de un cráter donde estuvo la ciudad de Excello, antes de que Dupree la destruyera. También descubre que la agente Sexton es en realidad la diosa Atenea de la mitología griega [40], la misma que protegió al clásico Hércules (y protege al Hércules del universo Marvel).

En la entrega final de la serie sobre la historia de Amadeus (el número 137) se descubre que Atenea hizo que Amadeus se salvara cuando Dupree envió a sus hombres para hacer explotar la casa de sus padres. Atenea (encarnada en una atractiva adolescente entonces) demoró a Amadeus, lo que lo hizo llegar tarde a la cena en su casa, y, por lo tanto, salvarse de morir en la explosión que mató a sus padres.

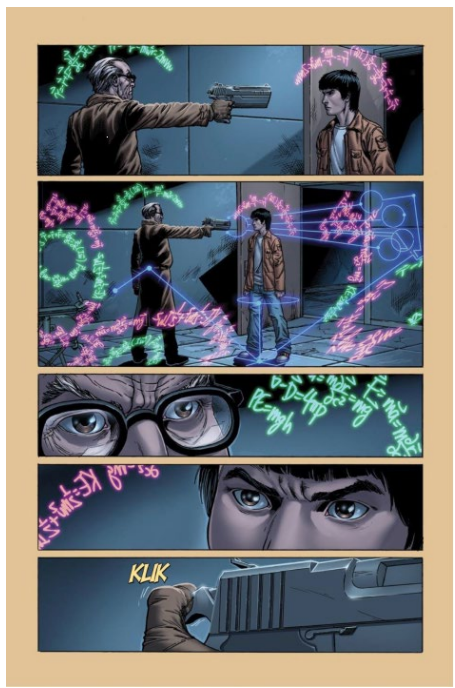

Figura 12. El enfrentamiento entre Amadeus Cho y Phytagoras Dupree en en *El Increíble Hércules* nro. 137.

En este capítulo se desarrolla un enfrentamiento entre Amadeus y Phytagoras Dupree que tiene lugar en la realidad real. No hay más cerebros de Boltzmann, pero se descubre que lo que le permitió a Amadeus ganar es que piensa en términos de ecuaciones. En la figura 12 se ve cómo este duelo de hiper-mentes es representado en la serie de historietas. En la página inicial de este número (la figura 13) puede verse cómo Amadeus relaciona ecuaciones con su situación. Sin embargo, las ecuaciones no tienen una relación clara con ninguno de los fenómenos físicos descritos anteriormente, ya que parecen ser ecuaciones asociadas con la mecánica newtoniana (por ejemplo, en la tercera viñeta de la figura 12 encontramos la ecuación de la energía potencial: PE=mgh).

Cuando las ecuaciones se relacionan efectivamente con la mecánica cuántica, no tienen nada que ver con las ecuaciones usadas en las especulaciones cosmológicas sobre los cerebros de Boltzmann (por ejemplo, en la figura 13 encontramos variaciones de las ecuaciones sobre el efecto Compton $\Delta\lambda = {}^{h}\!/_{mc}\,(1 - cos\Phi)$ [41]).

El hecho de que Amadeus tenga que explicarle a Atenea la física moderna, más específicamente las consecuencias de la mecánica cuántica que no son una "ilusión" refuerza los estereotipos de que las mujeres están vedadas de adentrarse en ese terreno (ni siquiera si son diosas). De hecho, en la larga discusión entre los antagonistas mentales nos enteramos de que Atenea se dio cuenta de que debía expandir el plantel de los héroes a los que apoyaba, porque los héroes de fuerza física (como Hércules) ya no podían prevalecer en el mundo moderno, regido por la ciencia. Necesitaba héroes de la mente, como Dupree (un experimento, que Atenea dice que le salió mal), o Amadeus Cho, quien eventualmente reemplazará a Hércules como héroe en las dos dimensiones posibles en las historietas de Marvel (mental y física) cuando se convierta en un nuevo tipo de Increíble Hulk.

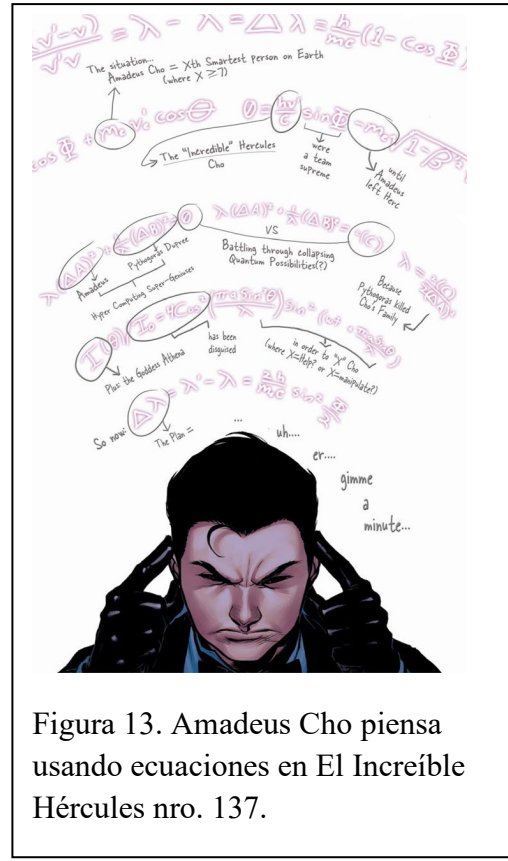

Figura 13. Amadeus Cho piensa usando ecuaciones en El Increíble Hércules nro. 137.

El uso de los cerebros de Boltzmann en esta serie confunde dos tipos de multiversos, en la clasificación de Tegmark [16], el cosmológico de la inflacionario eterna y el multiverso bifurcatorio de la interpretación de Everett. Aunque ambos han sido relacionados con los cerebros de Boltzmann, es más fácil o más probable que los cerebros surjan como una fluctuación en los primeros, y en ninguno de los dos hay conjeturas de que puedan ser "controlados" (o puedan controlar a los humanos u otros observadores evolutivos).

En resumen, la idea central es que la mente de estos personajes puede influenciar y manipular los multiversos tipo Everett en donde tratan de manipularse mentalmente. Los cerebros de Boltzmann son solamente instrumentos para que Dupree pueda crear las realidades paralelas donde encierra a Amadeus para derrotarlo. Los cerebros de Boltzmann son transitorios, porque solamente sirven para materializar realidades paralelas, comandados por el científico villano Dupree para impersonalizarse en esos mundos paralelos. Y pueden ser "desmantelados" si se observan en la forma correcta (usando las ecuaciones descifradas por Amadeus).

Para entender su uso en forma más clara, es necesario entender cómo funcionan los superhéroes y los supervillanos en el universo Marvel. Jason Bainbridge [42] ha argumentado (en una forma muy convincente en mi opinión) que tanto los superhéroes como los supervillanos transgreden las leyes del estado porque se consideran superiores a los humanos normales. Son super-humanos a la Nietzsche, que ejercen su voluntad de poder [43] porque las leyes humanas no permiten la justicia verdadera. En el caso de las historietas que hemos resumido, podemos ver que no sólo transgreden las leyes (humanas) del estado, sino también las de la física. El supervillano mental Pythagoras Dupree es capaz de manipular mentalmente las leyes de la mecánica cuántica, reclutando a cerebros de Boltzmann para crear multiversos mentales en donde lucha con Amadeus Cho para determinar cuál es la función de onda del universo que define/esconde la realidad "real". Amadeus responde a esa manipulación también en términos de la mecánica cuántica, encontrando las ecuaciones que revelan/desmantelan los trucos mentales del supervillano. Los cerebros de Boltzmann son poderosos instrumentos, y por fin, son transitorios.

## 6. La inevitable resurrección de los cerebros de Boltzmann: una novela con ecuaciones

Por último, voy a incluir información sobre una novela de mi autoría, que hace un uso extensivo del concepto de los cerebros de Boltzmann [44]. Como el nombre lo indica, la novela incluye 431 ecuaciones, no sólo de las ciencias exactas, sino también de otras disciplinas como la economía, la neurología y el cambio climático. En la tabla 1 se muestra cómo se incorporan ecuaciones de trabajos científicos que especulan sobre los cerebros de Boltzmann [45].

| La proporción de cerebros de Boltzmann (CB) sobre observadores normales (ON) en un multiverso | $\dfrac{\mathcal{N}^{CB}}{\mathcal{N}^{ON}} \approx \dfrac{\sum_j H_j^3 \mathcal{K}_j^{CB} s_j}{\sum_{i,k} n_{i,k}^{ON} \mathcal{K}_{i,k} s_k}$ |
|---|---|
| El factor de supresión de los cerebros de Boltzmann, donde $S_{CB}$ es la entropía y $F = M - T_{dS} S_{CB}$ es la energía libre del cerebro de Boltzmann | $\Gamma_{CB} \sim e^{-M/T_{dS}}\, e^{S_{CB}} \;=\; e^{-F/T_{dS}}$ |
| Estimación del número de cerebros de Boltzmann en un multiverso de volumen $e^{3H_A t}$ | $n_{CB} \sim \exp\left\{-N^{4/3}\dfrac{m_p}{m_e \alpha} + 3H_A t\right\}$ |

Tabla 1. Algunas ecuaciones relacionadas con los cerebros de Boltzmann encontradas en la literatura científica

En la novela, se explica el origen de estas entidades, y como puede verse en la siguiente cita (página 288), tienen mucha relación con las historietas:

*Una civilización andromedana, conocida como los vogons, había logrado erradicar las conciencias espontáneamente generadas en su entorno en lo que se conocía como la Tercera Guerra Termodinámica, que había obliterado una porción de un brazo externo de la galaxia y una de sus galaxias enanas satélites, donde vivían los kryptonianos, una civilización donde los cerebros de Boltzmann se habían establecido como parásitos, lo que les permitía extender significativamente su lapso de conciencia, ya que cuando aparecían de manera espontánea en la*

espuma cuántica del espacio tendían a desaparecer en unos pocos segundos. *Los vogons destruyeron Krypton, el planeta base de la civilización contaminada, y pusieron en una burbuja espacio-temporal cerrada los cerebros de Boltzmann que sobrevivieron para ser exiliados en una brana de un multiverso paralelo, aislado de la aglomeración biofílica. Sin embargo, los vogons no habían logrado esterilizar por completo su galaxia, ya que algunas naves espaciales con sobrevivientes kryptonianos contaminados habían escapado en dirección a la Vía Láctea, y habían liberado a esas entidades cerca del Sistema Solar. Las irresponsables transmisiones en busca de extraterrestres enviadas desde la Tierra les permitieron a los cerebros de Boltzmann identificar las características biológicas de los seres humanos, y su posible perpetuación en el planeta. Los eristeandos de Calipso, y los humanos en la Tierra habían sido las civilizaciones en donde esos seres se habían establecido como parásitos, siguiendo el mismo modelo de conquista de huéspedes biológicos que habían descubierto en Krypton.*

Aunque no se explica cuál es el mecanismo a través del cual contaminaron las mentes de los kryptonianos y los seres humanos, es esa la razón por la que pueden subsistir en el tiempo. También se explica cómo se reproducen (página 290):

*Al hablar con ellos un poco más y pedirles más explicaciones, se confundió aún más, porque en un sentido estricto, los cerebros de Boltzmann que ahora habitaban en la Tierra podían clasificarse como vida semi-evolutiva, y los únicos seres verdaderamente evolutivos serían aquellos que no habían sido contaminados. Le dijeron que su intuición era correcta, pero tenía también que entender como la vida evolutiva contaminada influenciaba la proliferación posterior de los parásitos. Si cinco o más de ellos se encontraban en una proximidad física de un radio de menos de un kilómetro, podían forzar la ecuación de estado del universo para crear una fluctuación similar a la que los había gestado, es decir, otro cerebro de Boltzmann, que copiaba la información que los otros ya habían obtenido y parasitaba nuevas conciencias*

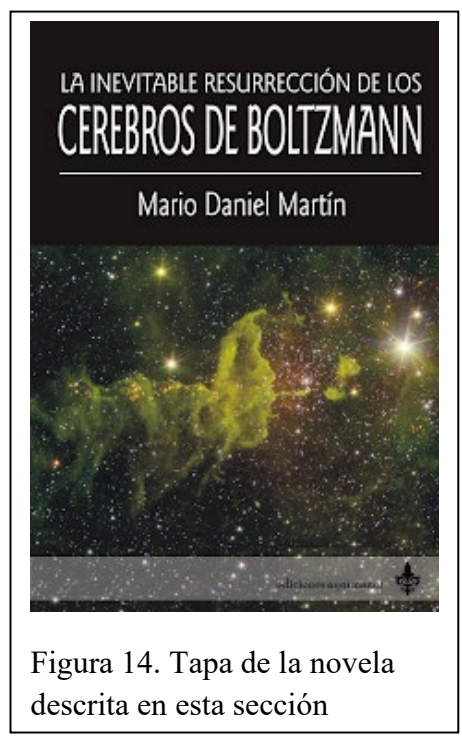

Figura 14. Tapa de la novela descrita en esta sección

*evolutivas del entorno. Eso había sucedido en la Tierra por primera vez en un pequeño pueblo llamado Smallville, que había hospedado a sobrevivientes kryptonianos, desde donde los cerebros de Boltzmann adaptados a contaminar seres humanos se habían extendido luego a Gotham City, y, de ahí, el resto del planeta.*

El mundo de las historietas (este caso el universo de DC Comics) ha sido usado como un contexto, en forma irónica claro está, en la novela. Sin arruinar las sorpresas de la trama para aquellos que decidan leerla, no puedo agregar mucho más. Pero los lectores interesados encontrarán muchas de las especulaciones sobre los cerebros de Boltzmann reflejadas en el argumento. En contraste con la situación en *El increíble Hércules*, estas entidades sí son personajes activos en la historia, ya que la contaminación de la glándula pineal de los líderes del planeta tierra por los cerebros de Boltzmann explica muchos de los problemas del Antropoceno [46] (página 291):

*En el Sistema Solar se había producido algo peculiar, esos cerebros de Boltzmann expulsados de la galaxia de Andrómeda por los vogons habían encontrado una forma de convertirse en parásitos de la vida evolutiva existente, y perpetuarse en el tiempo, produciendo la paradoja de que la una minoría de la humanidad, la que controlaba las decisiones, estaba compuesta de seres evolutivos que se comportaban como si no lo fueran.*

## Conclusión

En este trabajo he explorado los usos que se le ha dado a un concepto abstracto de la cosmología, los cerebros de Boltzmann, en diversos medios artísticos, focalizándome en cómo se usan en la ficción entidades que aparecen en medio del espacio sideral y luego desaparecen tan pronto como han nacido. He explorado, en particular, cómo el argumento de las obras de ficción justifica que puedan existir por un tiempo suficientemente largo para ser personajes viables. En las primeras dos obras (Futurama y Guardianes de la galaxia 2) no hay una justificación explícita. En las dos últimas, que mencionan explícitamente a los cerebros de Boltzmann, la justificación está basada en generosas interpretaciones de la mecánica cuántica.